\documentclass{aa}
\usepackage{graphics}
\def\a{{$\alpha$}}

\def\gsnr{{G~69.4+1.2}}
\newcommand{\h}{$^{\rm h}$}
\newcommand{\m}{$^{\rm m}$}
\newcommand{\s}{$^{\rm s}$}
\newcommand{\dd}{$\delta$}
\newcommand{\ha}{\rm H$\alpha$}
\newcommand{\hbeta}{\rm H$\beta$}
\newcommand{\HII}{\ion{H}{ii}}
\newcommand{\hnii}{{\rm H}$\alpha+[$\ion{N}{ii}$]$}
\newcommand{\nii}{$[$\ion{N}{ii}$]$}
\newcommand{\sii}{$[$\ion{S}{ii}$]$}

\newcommand{\oiii}{$[$\ion{O}{iii}$]$}
\newcommand{\snr}{\rm supernova remnant}

\newcommand{\et}{et al.}
\newcommand{\flux}{$10^{-17}$ erg s$^{-1}$ cm$^{-2}$ arcsec$^{-2}$}
\newcommand{\dens}{\rm cm$^{-3}$}
\newcommand{\vel}{\rm km s$^{-1}$}
\begin{document}

%
\title{Optical observations of the supernova remnant G 69.4+1.2}

\author{F. Mavromatakis\inst{1}
\and P. Boumis\inst{1}
\and E. V. Paleologou\inst{2}
}
\offprints{F. Mavromatakis,\email{fotis@physics.uoc.gr}}
\authorrunning{F. Mavromatakis et al.}
\titlerunning{Optical observations of \gsnr}
\institute{
University of Crete, Physics Department, P.O. Box 2208, 710 03 Heraklion, Crete, Greece 
\and Foundation for Research and Technology-Hellas, P.O. Box 1527, 711 10 Heraklion, 
Crete, Greece}
\date{Received 12 December 2001 / Accepted 14 March 2002}

\abstract{
We performed deep optical observations of the area of the new supernova 
remnant \gsnr\ in the emission lines of \oiii, \hnii\ and \sii. 
The low ionization images reveal diffuse and filamentary emission in the 
central and south, south--west areas of our field.  
Estimates of the \sii/\ha\ ratio suggest that the detected emission in these 
areas originates from shock heated gas, while the strong extended source in the
north must be an \HII\ region. 
The medium ionization image of \oiii\ shows a single filament close to the 
field center. Emission from \oiii\ is not detected elsewhere in the field 
but only in the north from LBN 069.96+01.35. 
Deep long--slit spectra taken at the position of the \oiii\ filament 
suggest shock velocities $\sim$120 \vel, while in other areas velocities 
around 50 \vel\ are expected. The sulfur lines ratio indicates electron
densities less than 120 \dens. The absolute \ha\ flux is $\sim$5 $\times$ 
\flux. The optical emission is very well correlated with the radio emission, 
especially in the south west. The soft X--ray emission detected in the ROSAT 
All-Sky survey shows a satisfactory degree of correlation with the 
optical data in the south--west suggesting their association.  
\keywords{ISM: general -- ISM: supernova remnants
-- ISM: individual objects: G 69.4+1.2}
}
\maketitle
\section{Introduction}
Most of the known supernova remnants have been discovered during radio 
surveys through their non--thermal synchrotron emission. 
A few remnants have been discovered by their optical or X--ray emission. 
The detection of X--ray emission in the direction of \object{G 69.7+1.0} 
was reported by Asaoka \et\ (\cite{asa96}). They found a 3\degr\ long  
arc in the east--west direction in the ROSAT All--Sky survey extending 
for 0\degr.8 in the south--north direction. 
The arc is convex to the north, its west end reaches the area of 
\object{CTB~80}, while the maximum surface brightness is found close to 
its south central areas. 
The Wolf-Rayet star WR 132 is found at a close angular distance but  
it is not expected to contribute to the morphology and energetics of the 
X--ray arc. WR 132 is considered to be at a distance of 4.4 kpc 
(Miller and Chu \cite{mil93}), much further than the estimated distance 
of $\sim$1 kpc to the arc (Asaoka \et\ \cite{asa96}, Lu \& Aschenbach
\cite{lu01}).  
Yoshita \et\ (\cite{yos00}) observed the area adjacent to G~69.7+1.0 
using the ROSAT and ASCA satellites. 
They detected diffuse thermal X--ray radiation with an average temperature 
of 0.42 keV attenuated by a neutral hydrogen column density of 
$\sim$1.6 $\times$ 10$^{21}$ cm$^{-2}$. This diffuse structure  extends for
$\sim$1\degr\ in the east--west direction and was designated as 
\object{AX~J2001+3235} or \object{\gsnr}. Yoshida \et\ (\cite{yos00}) proposed 
that the incomplete shells in radio and optical wavelengths around the 
X--ray emission are associated to the remnant.
The source AX~J2001+3235 is
actually the central part of the extended arc reported by Asaoka \et\
(\cite{asa96}). Lu \& Aschenbach (\cite{lu01}) fit the ROSAT pointed data 
with a two temperature Raymond--Smith model and obtained a neutral hydrogen 
density of $\sim$1.3 $\times$ 10$^{21}$ cm$^{-2}$ and temperatures of 
0.24 keV and 1.17 keV. They proposed that the whole 3\degr\ long X--ray
arc is \gsnr. Radio emission at 4850 MHz (Condon \et\ \cite{con94}) seems 
correlated with the X--ray emission, although its nature (thermal or
non--thermal) is not clear yet since radio spectral data are not 
available.  
\par
In order to improve our knowledge on this new supernova remnant 
and check whether the large optical shell is related to the X--ray
emission we performed 
deep optical observations of the area around G~69.7+1.0 in major optical 
emission lines. Information about the observations and the data 
reduction is given in Sect. 2. In Sect. 3 and 4 we present the results of 
our imaging observations and the results from the long--slit spectra. 
Finally, in Sect. 5 we discuss the physical properties of the remnant.
\section{Observations}
\subsection{Optical images}
The observations presented in this paper were performed with the 0.3 m 
Schmidt Cassegrain telescope at Skinakas Observatory. 
The field of \gsnr\ was observed in August 20 and 21, 2001. 
The observations were performed with a 1024 $\times$ 1024 Tektronix CCD which 
had a pixel size of 19 $\mu$m resulting in a 70\arcmin\ $\times$ 70\arcmin\ 
field of view and an image scale of 4\arcsec\ per pixel. 
A journal of the observations including exposure times and filter 
characteristics are given in Table~\ref{obs}.
The appropriate continua, also listed in Table~\ref{obs}, were subtracted 
from the low and medium ionization images. 
The final images in each filter are the average of the individual frames 
where appropriate, while the astrometric calculations utilized the HST 
Guide star catalogue. Two uncalibrated images in \hnii\ and \oiii\ were obtained
with the 1.3 m Ritchey--Cretien telescope at Skinakas Observatory and a 
1024 $\times$ 1024 SITe CCD. The exposure time of each frame was 1800 sec 
and the image scale is 0\arcsec.5 per pixel.
These frames were centered close to the position where the spectra were 
taken (see \S\ 2.2).  All coordinates quoted in this work refer to 
epoch 2000.
\par
We employed standard IRAF and MIDAS routines for the reduction of the data. 
Individual frames were bias subtracted and flat-field corrected using 
well exposed twilight flat-fields. The spectrophotometric standard stars 
HR5501, HR7596, HR7950, HR9087 and HR718 (Hamuy \et\ \cite{ham92},
\cite{ham94}) were used for absolute flux calibration.
%
  \begin {figure*}
   \resizebox{\hsize}{!}{\includegraphics{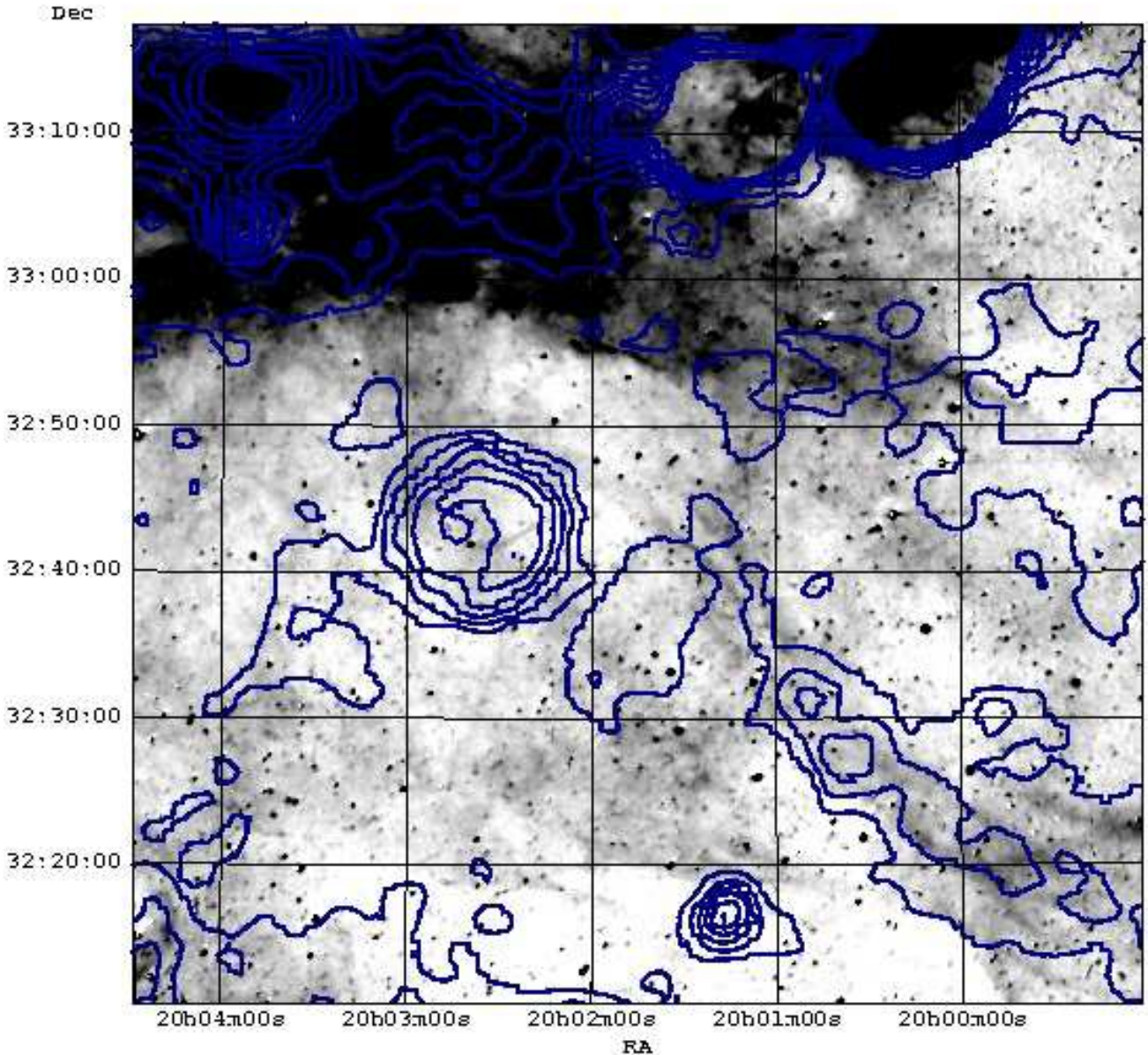}}
    \caption{ The 1\degr.1 field around \gsnr\ imaged with the  
    \hnii\ filter. The radio contours from 4850 MHz are overlaid to the 
     optical image. They scale linearly from 1.2 $\times$ 10$^{-4}$ Jy/beam to 
     9 $\times$ 10$^{-2}$ Jy/beam in steps of 11.24 $\times$ 10$^{-3}$ Jy/beam. 
     The optical image has been smoothed to suppress the residuals 
     from the imperfect continuum subtraction.
     Shadings run linearly from 0 to 50 $\times$ \flux. 
     The extended bright source to the north is LBN 069.96+01.35.} 
     \label{fig01}
  \end{figure*}

\subsection{Optical spectra}
Long--slit spectra were obtained on July 28, 2001  using the 
1.3 m Ritchey--Cretien telescope at Skinakas Observatory. 
The data were taken with a 1300 line mm$^{-1}$ grating 
and a 800 $\times$ 2000 SITe CCD covering the range of 4750 \AA\ -- 6815 \AA.
The slit had a width of 7\farcs7 and, in all cases, was oriented
in the south-north direction. The slit center was located at 
$\alpha$ = 20\h01\m33\s, $\delta$ = 32\degr43\arcmin30\arcsec\ and two
exposures were obtained, each of 3900 sec. 
The spectrophotometric standard stars HR5501, HR7596, HR7950, HR9087, and 
HR718 (Hamuy \et\ \cite{ham92}, \cite{ham94}) were observed in order to 
calibrate the spectra.  
\begin{table*}
      \caption[]{Journal of the Observations}
         \label{obs}
\begin{flushleft}
\begin{tabular}{lllllll}
            \noalign{\smallskip}
\hline
   &   $\lambda_{\rm C}$   & FWHM   &  & &
 &   \cr 
Filter &    (\AA)  &   (\AA)  & Telescope & Date (UT) &
Frames$^{\rm a}$ & Exposure$^{\rm b}$ (s) \cr 
\hline
\hnii   &    6560                   & 75          & 0.3 m   & 20--21 Aug., 2001
& 2 & 4800    \cr
\sii    &    6708                   & 20          & 0.3 m   & 20 Aug., 2001 
& 2 & 4800   \cr
\oiii   &     5005                  & 28          & 0.3 m   & 21 Aug., 2001
& 3 & 7200 \cr
Cont red  &  6096                  & 134         & 0.3 m   & 20--21 Aug., 2001
& 8 & 1440 \cr
Cont blue  & 5470                  & 230         & 0.3 m   & 21 Aug., 2001     
& 4 & 720 \cr
\hline
\hnii  & 6570                       & 75          & 1.3 m   & 11 Aug., 2001 
& 1 & 1800 \cr
\oiii  & 5014                       & 28          & 1.3 m   & 11 Aug., 2001 
& 1 & 1800 \cr  
\hline
\end{tabular}
\end{flushleft}
${\rm ^a}$ Number of individual frames \\\
${\rm ^b}$ Total exposure times in sec \\\
\end{table*}
\section{Imaging of \gsnr}
\subsection{The \hnii\ and \sii\ line images}
\begin {figure}
   \resizebox{\hsize}{!}{\includegraphics{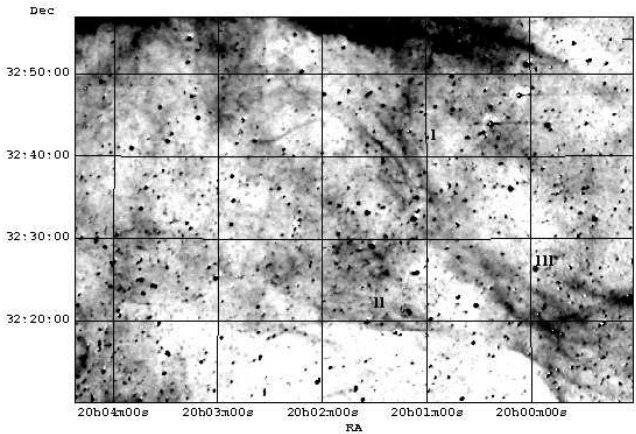}}
    \caption{ In this figure we show only the south part of our field
     as is seen in the \sii\ filter due to the strong \HII\ region to 
     the north. 
     The image has been smoothed to suppress the residuals 
     from the imperfect continuum subtraction and the shadings run 
     linearly from 0 to 25 $\times$ \flux.} 
     \label{fig02}
  \end{figure}

Our field  (Fig. \ref{fig01}) includes the known \snr\ G~69.7+1.0 as well 
as the strong and extended \HII\ region LBN 069.96+01.35 (Lynds \cite{lyn65}). 
We detect diffuse emission and a filamentary structure at the area 
where G~69.7+1.0 lies, however, it is not clear if this emission is associated 
with this remnant or \gsnr. 
More emission line structures are detected in the south areas of our field. 
The pulsar \object{PSR 2002+3217} is within the field of view but  
no emission is detected from this source. In Fig. \ref{fig01} we also 
show the radio emission at 4850 MHz (Condon \et\ \cite{con94}) in contours
scaling linearly from 1.2 $\times$ 10$^{-4}$ Jy/beam to 
9 $\times$ 10$^{-2}$ Jy/beam. The radio data were retrieved 
from {\it SkyView} maintained by NASA/HEASARC.
The \hnii\ image shows filamentary emission extending from the 
field center to the south west. Two filaments running parallel for 
$\sim$15\arcmin\ are present around \a\ $\simeq$ 20\h01\m\ and 
\dd\ $\simeq$ 32\degr40\arcmin.
In addition, more emission features are found to the south 
between \dd\ $\simeq$ 32\degr20\arcmin\ and 32\degr30\arcmin. 
The same morphological characteristics observed in the \hnii\ image are also 
seen in the \sii\ image (Fig. \ref{fig02}). However, some differences do exist 
among the two images. The east one of the two filaments 
(\a\ $\simeq$ 20\h01\m15\s, \dd\ $\simeq$ 32\degr40\arcmin) mentioned above does 
not appear continuous in the \sii\ image but gaps in intensity along the 
filament are seen.  
The emission measured in 
the \sii\ image itself is quite strong relative to the \hnii\ emission.
In Table~\ref{fluxes}, we list typical fluxes measured in several locations 
(areas I, II, III in Fig. \ref{fig02}) within the field of \gsnr\ including the 
\HII\ region LBN 069.96+01.35 (LBN 159). 
The \nii 6548, 6584 \AA\ lines and the \ha\ line contribute equally in the 
\hnii\ filter and the corresponding fluxes in Table~\ref{fluxes} include all
these lines. However, the sulfur lines at 6716 \AA\ and 6731 \AA\ are not
transmitted equally in the \sii\ filter because its bandpass is narrower 
than that of the \hnii\ filter. Thus, the sulfur fluxes quoted in
Table~\ref{fluxes} include 100\% of the 6716 \AA\ line and 18\% of the 6731 \AA\
line.  
Since the images are flux calibrated, the \hnii\ and \sii\  frames can be used 
to study the nature of the detected emission. 
We estimate \sii/\ha\ values of $\sim$0.9 in the south, south--west suggesting
emission from shock heated gas, while in several locations of LBN 069.96+01.35
we estimate values of $\sim$0.4 indicating \HII\ emission. 
In order to obtain these quantitative estimates we proceeded as follows. 
Since our \sii\ filter does not measure the total sulfur flux we assumed a ratio
of I(6716 \AA)/I(6731 \AA) of $\sim$1.4, appropriate for evolved remnants, and
the filter transmissivities (100\% at 6716 \AA\ and 18\% at 6731 \AA) to
estimate the total sulfur flux. In addition, we assumed that the \ha\ flux 
is typically $\sim$90\% or greater of the \nii 6548, 6584 \AA\  flux for evolved 
remnants while for \HII\ regions this percentage is less than $\sim$50\%--60\%. 
Under these assumptions we estimate that 
\sii/\ha\ $\sim$2.7([\sii/\hnii)$_{\rm observed}$.
%
%
  \begin{table}
      \caption[]{Typically measured fluxes}
         \label{fluxes}
\begin{flushleft}
\begin{tabular}{lllll}
            \hline
            \noalign{\smallskip}
     & Area I  & Area II & Area III & LBN 159  \cr	
\hline
\hnii\ 	& 40	& 30  	& 40	&   90    \cr	
\hline
\sii\ 	& 12	& 14	& 16   &   20 	\cr	
\hline
\oiii\ 	& 7	& --	& --   &  10 \cr	
\hline
\end{tabular}
\end{flushleft}
${\rm }$ Fluxes are median values over several 40\arcsec $\times$
40\arcsec\ \\ boxes as measured in the corresponding filters\\
${\rm }$ and in units of \flux \\\
${\rm }$ See \S 3.1 for more details
\end{table}
  \begin {figure}
   \resizebox{\hsize}{!}{\includegraphics{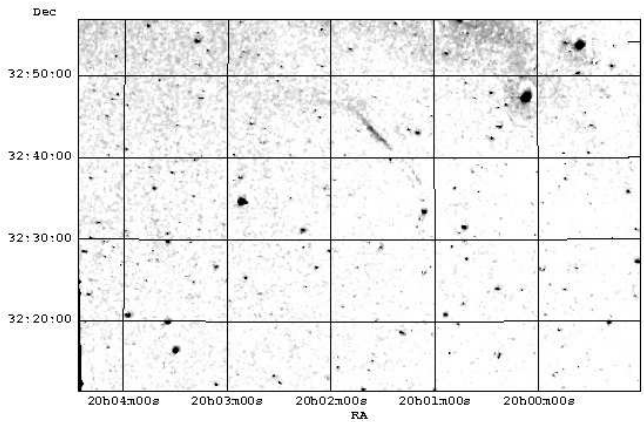}}
    \caption{ The area around \gsnr\ imaged with the medium ionization 
     line of \oiii 5007\AA. The image has been smoothed to suppress the 
     residuals from the imperfect continuum subtraction and the shadings run 
     linearly from 0 to 15 $\times$ \flux.} 
     \label{fig03}
  \end{figure}
\subsection{The \oiii\ image}
The morphology of the medium ionization line of \oiii\ (Fig. \ref{fig03})  
is markedly different from that seen in the \hnii\ and \sii\ images. 
Typical fluxes measured in the \oiii\ calibrated image are also given in 
Table~\ref{fluxes}. 
A filament is detected close to the field center around at 
\a\ $\simeq$ 20\h01\m33\s\ and \dd\ $\simeq$ 32\degr43\arcmin30\arcsec. 
This filament is not continuous over its full extent but there is a gap of 
$\sim$4\arcmin. This probably reflects inhomogeneities 
in the interstellar ``clouds'' resulting in strong variations of the shock 
velocity upon which the \oiii\ flux depends crucially (Cox \& Raymond 
\cite{cox85}). An 8\arcmin.5 $\times$ 8\arcmin.5 square area has been 
imaged in better angular
resolution (0\arcsec.5 per pixel) with the 1.3 m telescope through the 
\hnii\ and \oiii\ interference filters (Fig. \ref{fig04}). The arrows in 
Fig. \ref{fig04} indicate the location of the slit center. 
A wider field was observed by Miller and Chu (\cite{mil93}) during their 
study of the nearby Wolf-Rayet star WR~132. 
Our images show very narrow filamentary structures as well as diffuse
emission. The \hnii\ image shows a $\sim$1\arcmin\ wide lane free of emission  
to the west of the observed filament but across this gap diffuse emission 
is present. The projected angular width of the filament seen in the low 
ionization image is $\sim$16\arcsec, while the width of the filament seen 
in the medium ionization line of \oiii\ is only $\sim$5\arcsec. 
At the very south--west edge of the high resolution image \oiii\ filamentary 
emission is again detected (see also Fig. \ref{fig03}). 
This emission is enclosed by the dashed ellipse 
in Fig. \ref{fig04}. The failure to detect \oiii\ emission where the \hnii\ 
emission is strong suggests that these lines are anti-correlated.  
No \oiii\ emission is detected in other areas of the remnant, including 
those areas where diffuse or filamentary \hnii\ or \sii\ emission is observed. 
\section{The long--slit spectra from \gsnr}
\begin {figure}
   \resizebox{\hsize}{!}{\includegraphics{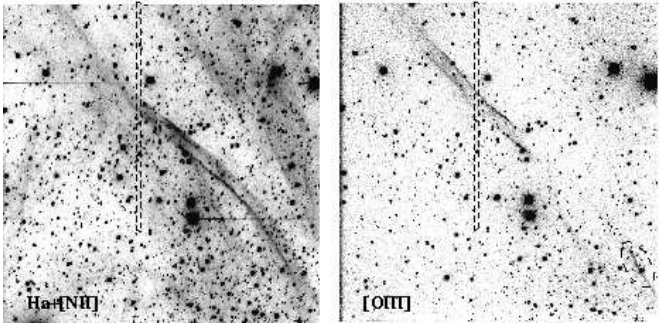}}
    \caption{High resolution images in \hnii\ (left) and \oiii\ (right) 
     of the area where \oiii\  emission is prominent. 
     The field is 8\arcmin.5 $\times$ 8\arcmin.5 and the dashed rectangles 
     show the projection of the slit on the sky. 
     The dashed ellipse in the lower right of the \oiii\ image encloses faint 
     filamentary emission.}
     \label{fig04}
  \end{figure}

The deep low resolution spectra were taken at the location of the 
bright \oiii\ filament. As is evident from Fig. \ref{fig04} the 
corresponding \ha\ emission is expected to be weaker. 
The spectra clearly demonstrate the fact that the observed emission must 
originate from shock heated gas (Table~\ref{sfluxes}). 
Note that the fluxes quoted in Table~\ref{sfluxes} 
are expected to differ from those given in Table~\ref{fluxes} since 
the latter are averages over a sufficient number of pixels at different 
locations around the field observed, while the former refer to the very 
restricted position of the slit. Note also that the slit was placed at the
position of bright \oiii\ emission and not \ha\ emission. The \ha\  emission at
the slit location is much weaker than at neighbour positions as is evident from 
Fig.~\ref{fig04}.
The ratio of the sulfur lines approaches the low density 
limit indicating low electron densities (e.g. Osterbrock \cite{ost89}). 
However, given the errors on the individual sulfur fluxes, we estimate 
that electron densities less than $\sim$120 \dens\ (1$\sigma$) or 200 \dens\ 
(2$\sigma$) are allowed.
The spectra display strong \oiii\ emission relative to \hbeta\ suggesting shock 
velocities greater than $\sim$100 \vel\ (Cox \& Raymond ~\cite{cox85}, 
Hartigan \et\ ~\cite{har87}). We note here that the signal to noise ratios quoted 
in Table~\ref{sfluxes} do not include calibration errors which are $\sim$10\%. 
\section{Discussion}
\begin {figure*}
   \resizebox{\hsize}{!}{\includegraphics{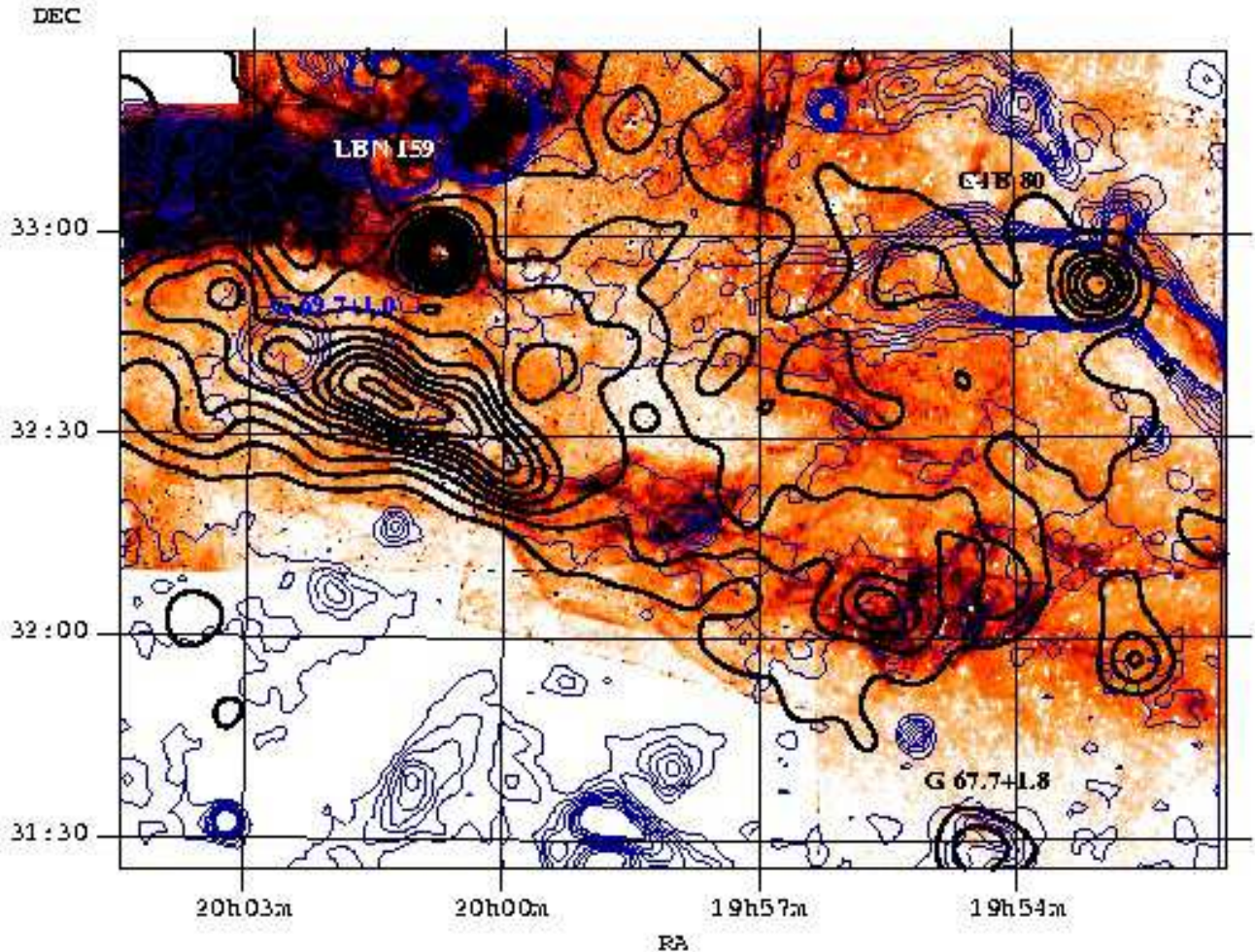}}
    \caption{ A complex field seen through the \hnii\ filter around \gsnr. 
    The remnants CTB~80, G~69.7+1.0, G~67.7+1.8 are within the field of view.
    Several structures are also present in the south, south--west. 
    The radio contours at 4850 MHz are overlaid on the 
    optical data in a linear scale from 0.002 Jy/beam to 0.1 Jy/beam 
    (thin contours). The thick contours represent the soft X--ray emission 
    seen in the ROSAT All--Sky survey. A color version of this plot is 
    available on the electronic edition of this work.
     } 
    \label{fig05}
   \end{figure*}

We performed the first deep optical observations of the immediate vicinity 
of the new \snr\ \gsnr\ in major emission lines. The CCD images obtained reveal 
diffuse and filamentary emission, in the low ionization lines, in the south, 
south--west areas. The survey of Parker \et\ (\cite{par79}) was examined  
but we were not able to identify the structures that we have detected in this
work. This is probably due to the lower resolution and sensitivity of this 
survey. The supernova remnant \object{G 69.7+1.0} lies within our field but we did
not find any strong evidence to claim the detection of optical emission from
this remnant. The flux calibrated images allow us to conclude that the 
bright extended source LBN 069.96+01.35 is  an \HII\ region, 
while the structures
detected in the central and south areas of our field result from emission 
of shock heated gas since their \sii\ emission is strong relative to \ha. 
Note here that LBN 069.96+01.35 is found in the north part of the large 
optical shell reported by Yoshida \et\ (\cite{yos00}) to be associated with the
X--ray emission.
In order to understand better the properties of \gsnr\ we combined published 
data on \object{G 67.7+1.8} and \object{CTB 80} (Mavromatakis \et\ \cite{mav01a},
\cite{mav01b}) with the current data to create a 2\degr.8 $\times$ 1\degr.8 
mosaic in \hnii\ (Fig. \ref{fig05}) . In this figure we overlaid the 4850 MHz
radio data (thin contours; Condon \et\ \cite{con94}) as well as public ROSAT
All--Sky survey data (thick contours; see also Lu \& Aschenbach \cite{lu01}). 
It is clear from the wide--field optical image that the filamentary structures
detected in the south--west (Figs. \ref{fig01}, \ref{fig02}) extend
further to the west, up to \a\ $\simeq$ 19\h54\m.
The radio emission, although weak, is very well correlated with the optical line
emission all the way from \a\ $\simeq$ 19\h54\m\ to 20\h01\m\ (Figs. \ref{fig01},
\ref{fig05}). Even though good quality radio data at other wavelengths are not 
available, the nature of the radio emission is probably  
non--thermal given the strong sulfur line emission seen in the optical data. 
Thus, we propose that the detected line emission is the optical counterpart 
of the radio emission seen at 4850 MHz. The question that naturally 
arises is whether the optical or radio emission is related to the 
soft X--ray emission. Clearly, the possibility that they are not related and 
simply overlap cannot be excluded. 
In this case we would have to propose the detection of a candidate remnant, 
different from \gsnr, based on our observational evidence. 
However, the nature of the optical emission 
and possibly that of the radio, the common orientation of the observed 
emission (south--west to north--east) and the fact that the outer soft X--ray 
contours roughly trace the south boundaries of the optical and radio emission
would favor the association of the optical radiation with the emission detected
in the ROSAT All--Sky survey. 
\par
The image in the medium ionization line of \oiii\ shows a filament close to
the field center, while no other \oiii\ emission is detected except from the 
\HII\ region. We note that this remnant belongs to the class of remnants 
which show prominent \oiii\ emission only in a specific portion of 
their emission like \object{CTB 1} (Fesen \et\ \cite{fes97}), \object{CTB 80},
\object{G 114.3+0.3} (Mavromatakis \et\ \cite{mav01b}, \cite{mav02}) and 
\object{G 17.4-2.3} (Boumis \et\ \cite{bou02}).
The failure to detect \oiii\ line emission in several locations
suggests that the velocity of the shock propagating into the interstellar
``clouds'' is less than $\sim$100 \vel, while the detection of strong \oiii\ 
emission relative to \ha\ at a specific position implies shock velocities 
in excess of $\sim$120 \vel\ (Hartigan \et\ \cite{har87}) assuming complete 
recombination zones. The relation 
\begin{equation}
{\rm n_{[\ion{S}{ii}]} \simeq\ 45\ n_c \times V_{s,100}^2},
\end{equation}
given by Fesen \& Kirshner (\cite{fes80}) can be used to limit the preshock 
cloud density n$_{\rm c}$. The electron density derived from the sulfur 
line ratio is denoted by ${\rm n_{[\ion{S}{ii}]}}$ and V$_{\rm s,100}$ is 
the shock velocity into the clouds in units of 100 \vel. 
This relation allows us to set an upper limit on the preshock cloud density 
of $\sim$2 \dens. Here we assumed a shock velocity of 120 \vel\ and made use of the 
upper limit of 120 \dens\ on the electron density (Table \ref{sfluxes}). 
However, the relatively high \oiii\ to \hbeta\ ratio in 
connection with the different morphologies seen in Fig. \ref{fig04} could 
suggest that incomplete recombination zones are present in that area. 
Adopting the calculations of Raymond \et\ (\cite{ray88}), a distance of 
$\sim$1 kpc (Asaoka \et\ \cite{asa96}, Lu \& Aschenbach \cite{lu01}) 
and an edge--on geometry, we estimate an average filament density 
of $\sim$4 \dens. We note here that in this case a shock velocity of 
$\sim$80 \vel\ would be sufficient to produce the strong \oiii\ emission. 
However, it is not clear if this velocity could also produce the prominent 
sulfur emission seen in the images. 
\par
We estimate from the flux calibrated images that the total sulfur emission at 
6716 \AA\ and 6731 \AA\ is comparable in strength to the total nitrogen 
emission at 6548 \AA\ and 6584 \AA. 
In addition, emission from \oiii\ is practically absent
since it is only present in a very limited area compared to the field observed 
(Fig. \ref{fig03}). Examination of shock models with full recombination zones 
shows that these conditions are satisfied only for low shock velocities 
around 40--60 \vel\ (Hartigan \et\ \cite{har87}). Clearly, such velocities 
can account for the gross absence of the \oiii\ emission. 
In order to obtain more quantitative estimates we will also use the relation 
\begin{equation}
{\rm E_{51}} = 2 \times 10^{-5} \beta^{-1}\ {\rm n_c}\ V_{\rm s,100}^2 \ 
{\rm r_{s}}^3,  
\end{equation}
given by McKee \& Cowie (\cite{mck75}). 
The factor $\beta$ is of the order of 
1--2, ${\rm E_{51}}$ is the explosion energy in units of 10$^{51}$ erg, 
and  r$_{\rm s}$ the radius of the remnant in pc.
Assuming r$_{\rm s}$ = 28 (Lu \& Aschenbach \cite{lu01}), 
V$_{\rm s,100}$ $\simeq$ 0.5,  n$_{\rm c}$ $\simeq$ 1 and $\beta$ $\simeq$ 1, 
we estimate an explosion energy of 0.1 which is not far from the value of 0.5 
given by Lu \& Aschenbach (\cite{lu01}) considering the estimates of the 
physical quantities and the assumptions made in the equations involved. 
Furthermore, we can check if pressure equilibrium exists between the cloud and
intercloud regions. Using as before the same values for the shock velocity and 
precloud density and a velocity of the main shock front of $\sim$700 \vel\ 
(Lu \& Aschenbach \cite{lu01}), we calculate an intercloud density of 0.01 \dens. 
This value is very close to that of 0.02 \dens\ given by Lu \& Aschenbach 
(\cite{lu01}) based on the Sedov--Taylor solution of the X--ray data.
These simple calculations show that the optical data are compatible with basic 
\snr\ parameters derived from the X--ray data. Nevertheless, new X--ray
observations are needed with a better spatial resolution in order to perform a
detailed comparison with the optical and radio data. Radio observations are also
needed to determine the actual nature of the radio emission in the south,
south--west as well as in the north.
\section{Conclusions}
The area around the new supernova remnant \gsnr\ was observed in major 
optical lines. New diffuse and filamentary structures were detected in 
the south, south--west. The flux calibrated images suggest that the 
majority of the structures result from emission of shock heated 
gas. Variations in the shock velocity are also observed, while the 
long--slit spectra taken at the position of prominent \oiii\ emission 
suggest low electron densities. 
The optical data seem correlated with the radio data at 4850 MHz 
and there is also good agreement with the X--ray data although the spatial 
resolutions are quite different. It is very likely that the 
optical emission is related to the radio emission and to the extended 
X--ray arc. 
\begin{acknowledgements}
\end{acknowledgements}
We would like to thank the referee F. Winkler for his comments 
which helped to clarify the scope of this paper. 
We would also like to thank B. Aschenbach for providing us a preprint 
on \gsnr\ (Lu \& Aschenbach \cite{lu01}). 
Skinakas Observatory is a collaborative project of the University of
Crete, the Foundation for Research and Technology-Hellas and
the Max-Planck-Institut f\"ur Extraterrestrische Physik.
This research has made use of data obtained through the High Energy 
Astrophysics Science Archive Research Center Online Service, 
provided by the NASA/Goddard Space Flight Center.
 \begin{table}
        \caption[]{Relative line fluxes}
         \label{sfluxes}
         \begin{flushleft}
         \begin{tabular}{lllllll}
     \hline
 \noalign{\smallskip}
Line (\AA) & F$^{\rm a,b}$  \cr
\hline
4861 \hbeta\   & 13 (3)$^{\rm c}$    \cr
\hline
4959 \oiii\  & 29 (7)    \cr
\hline
5007 \oiii\  & 106 (23)    \cr
\hline  
6548 \nii\    & 47 (21)   \cr
\hline
6563 \ha\     & 100 (41) \cr
\hline
6584 \nii\    & 151 (50) \cr
\hline
6716 \sii\    & 98 (34)  \cr
\hline
6731 \sii\    & 72 (27)  \cr
\hline
\hline
Absolute \ha\ flux$^{\rm d}$ & 5.3 (10)     \cr
\hline
\ha /\hbeta\   & 7.8 (3)  \cr
\hline
\sii/\ha\ 	& 1.7 (30) \cr
\hline 
F(6716)/F(6731)	& 1.4 (21)\cr
\hline 
\end{tabular}
\end{flushleft}
${\rm ^a}$ Uncorrected for interstellar extinction \\
${\rm ^b}$ Listed fluxes are a signal to noise weighted
average of the individual fluxes\\
${\rm ^c}$ Numbers in parentheses represent the signal to noise ratio 
of the quoted fluxes\\
$^{\rm d}$ In units of \flux\\\
${\rm }$ All fluxes normalized to F(\ha)=100
\end{table}
\vfill\eject

\end{document}